\documentclass[apl,twocolumn,floatfix,footinbib,showpacs,superscriptaddress]{revtex4-2} %IZ 4-2
\usepackage{graphicx} 
\usepackage{amsfonts,amsmath,amssymb}
\usepackage{amsthm}
\usepackage{dcolumn}
\usepackage{dsfont,bm}
\usepackage[colorlinks=true,linkcolor=blue,pagecolor=blue,filecolor=blue,menucolor=blue,urlcolor=blue,citecolor=blue,anchorcolor=blue]{hyperref}% add hypertext capabilities
\usepackage{color}
\usepackage{soul} 
\usepackage{amsbsy}
\usepackage{float}
\usepackage{hyperref}

\begin{document}

\title{Tunable planar Josephson junctions driven by time-dependent spin-orbit coupling}

\author{David Monroe }
\affiliation{University at Buffalo, State University of New York, Buffalo, New York 14260-1500, USA}
\author{Mohammad Alidoust}
\affiliation{Department of Physics, Norwegian University of Science and Technology, N-7491 Trondheim, Norway}
\author{Igor \v{Z}uti\'{c}}
\affiliation{University at Buffalo, State University of New York, Buffalo, New York 14260-1500, USA} 
\hfill
\date{Received 18 April 2022; revised 10 July 2022; accepted 2 August 2022; published 15 September 2022}

\begin{abstract}
The integration of conventional superconductors with common III–V semiconductors provides a versatile platform to implement tunable Josephson junctions (JJs) and their applications. 
We propose that with gate-controlled time-dependent spin-orbit coupling, it is possible to strongly modify the current-phase relations and Josephson energy and provide a mechanism to drive the JJ dynamics, even in the absence of any bias current.  We show that the transition between
stable phases is realized with a simple linear change in the strength of the spin-orbit coupling, while the transition rate can exceed the gate-induced electric field gigahertz changes 
by an order of magnitude. The resulting 
interplay between the constant effective magnetic field and changing spin-orbit coupling has direct implications
for superconducting spintronics, the control of Majorana bound 
states, and emerging qubits. We argue that topological superconductivity, 
sought for fault-tolerant quantum computing, offers simpler applications in superconducting electronics and spintronics. \\ \\
DOI: \href{http://dx.doi.org/10.1103/PhysRevApplied.18.L031001}{10.1103/PhysRevApplied.18.L031001}
\end{abstract}

\maketitle

In the push to implement beyond-CMOS applications, Josephson junctions (JJs) have found their broad use due to their
high-speed switching, low-power dissipation, and intrinsic nonlinearities\cite{Tafuri:2019,Siegel:2012}. In addition to the well-established role of JJs
as the key elements for superconducting electronics and superconducting qubits\cite{Tafuri:2019,Siegel:2012,Krantz2019:APR,Likharev1991:IEEETAS,%
Holmes2013:IEEETAS,Giazotto2010:NP},
there is a growing interest in tailoring their spin-dependent properties to enable dissipationless spin currents, cryogenic memory\cite{Eschrig2015:RPP,Khaire2010:PRL,Banerjee2014:NC,Gingrich2016:NP,Birge2019:IEEEML,Mazanik2020:PRA,Linder2015:NP,Han2020:NM}, 
and fault-tolerant quantum computing\cite{Fu2008:PRL,Aasen2016:PRX,Laubscher2021:JAP,Gungordu2022:P,Rokhinson2012:NP,Fornieri2019:N,Ren2019:N,Desjardins2019:NM}.
The role of spin-orbit coupling (SOC) has been extensively studied in the normal-state properties and recognized
for its importance in spintronics\cite{Zutic2004:RMP,Fabian2007:APS,Bercioux2015:RPP}.
However, the superconducting analogs of the SOC-related effects remain to be understood. They might even be important when their normal-state counterparts 
are negligibly  small\cite{Hogl2015:PRL,Costa2017:PRB,Lv2018:PRB,Martinez2020:PRA,Vezin2020:PRB,Gonzalez-Ruano2020:PRB,Cai2021:NC}. Motivated
by the recent progress in gate-controlled SOC in planar JJs based on a two-dimensional electron gas (2DEG)\cite{Mayer2020:NC,Dartiailh2021:PRL,Mayer2020:AEM}, 
we reveal how time-dependent SOC tunes many of their key properties and offers an unexplored mechanism to drive JJs.

A common description of a JJ circuit, 
is given by a Josephson element, resistor, and capacitor connected in parallel, using the resistively and capacitively shunted junction (RSCJ)\cite{Tafuri:2019} model.
The bias 
current through 
the junction, $i$, is the sum of the supercurrent and the quasiparticle current flowing in  
the resistor 
and capacitor. The supercurrent is usually assumed as $I(\varphi)=I_c \sin(\varphi+\varphi_0)$, where $I_c$ is the maximum supercurrent,
$\varphi$ is the phase difference between the superconducting regions, and
the anomalous phase, $\varphi_0\neq 0, \pi$,
arises from the broken time-reversal and inversion symmetries\cite{Reynoso2008:PRL,Buzdin2008:PRL,Konschelle2009:PRL,Sickinger2012:PRL,Strambini2020:NN,Shukrinov2022:PU,Soloviev2021:PRA}.

\begin{figure}
\vspace{-1.4cm}
\includegraphics[width=1.6\columnwidth]{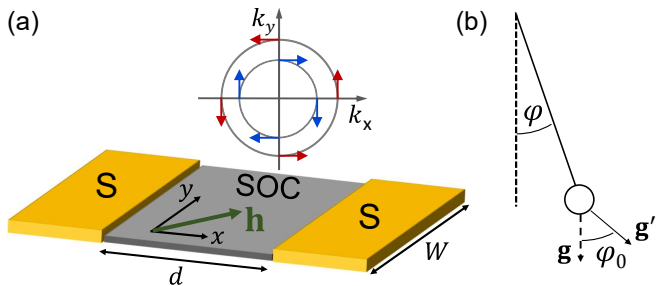}
\vspace{-5.7cm}
\caption{(a) A schematic of the Josephson junction (JJ). Two $s$-wave superconductors ($S$), are separated by the middle region 
which hosts the Rashba spin-orbit coupling (SOC), with depicted ${\bf k}$-space spin-orbit fields, and an effective Zeeman field, ${\bf h}$.
(b) A mechanical pendulum model of the JJ. The displacement angle $\varphi$ is analogous to the superconducting phase difference and ${\bf g}$
is the gravitational acceleration for vanishing SOC and ${\bf h}$. The pendulum is driven by changing the effective ${\bf g'}$,
an interplay between ${\bf h}$ and time-dependent SOC.  This yields a tunable current-phase relation and
an anomalous phase, $\varphi_0$, equivalent to the equilibrium of the displaced pendulum.} 
\label{fig:f1}
\end{figure}

For a ballistic JJ depicted in Fig.~\ref{fig:f1}(a),  
the interplay between SOC and the effective Zeeman field ${\bf h}$, 
yields  
a more complex current-phase relation (CPR) than $I(\varphi)$ given above, such that for a generalized
RSCJ model  
\begin{equation}
d^2\varphi/d\tau^2 + (d\varphi /d\tau)/\sqrt{\beta_c} + I(\varphi, \mu, {\bf h}, \alpha)/I_c = i/I_c,
\label{eq:pendulum}
\end{equation}
 where  
$\tau = \omega_p t$ is a dimensionless time, expressed using the JJ plasma frequency,  $\omega_p = \sqrt{2\pi I_c / \Phi_0 C}$, 
$\Phi_0=h/2e$ is the magnetic flux quantum, and $C$ is the capacitance. The damping of this nonlinear oscillator is characterized by the Stewart-McCumber parameter, 
$\beta_c = 2\pi I_c C R^2 / \Phi_0$, where  $R$ is the resistance\cite{Stewart1968:APL,McCumber1968:JAP} and $Q=\sqrt{\beta_c}$ is the quality factor.
The generalized CPR can be modified by the chemical potential $\mu$, and  ${\bf h}$, arising from the applied magnetic field
or magnetic proximity effect\cite{Zutic2019:MT}. Since 
$h_z$ does not induce $\varphi_0$\cite{Alidoust2021:PRB,Alidoust2020:PRB} and  
only produces CPR reversals, we focus on $h_z=0$ [Fig.~~\ref{fig:f1}(a)]. 
The CPR can also be tuned 
by the Rashba SOC, illustrated in Fig.~\ref{fig:f1}(a),  which is parametrized by its strength $\alpha$, 
in the Hamiltonian, $H_\mathrm{so} =  \alpha (\bf\sigma \times {\bf p}) \cdot {\bf \hat{z}}$. 
Here, $\bf\sigma$ is the Pauli matrix vector, and $\bf p$ is the in-plane momentum,
for 2DEG with the inversion symmetry broken along the $z$-direction\cite{Bychkov1984:PZETF}. 

While quasistatic gate-tunable changes in SOC and $\varphi_0$ have been 
demonstrated in 2DEG-based JJs\cite{Mayer2020:NC,Dartiailh2021:PRL}, the implications of dynamically tuned SOC on the CPR remain unexplored. 
For a conventional CPR without any $\varphi_0$, Eq.~(\ref{eq:pendulum}) has a mechanical analog with 
a driven and damped pendulum, in which $\varphi$ becomes the displacement angle\cite{Stewart1968:APL,McCumber1968:JAP}. 
A JJ driven by $i$  is equivalent to the pendulum displaced by an external torque from its stable equilibrium, determined by the gravitational acceleration 
$\bf g$, while $\omega_p$ determines the oscillation frequency around a stable equilibrium point\cite{Tafuri:2019}. 

Instead of using $i$, Fig.~\ref{fig:f1}(b) suggests an entirely different way to drive the pendulum: By changing the orientation of the 
effective $\bf g'$ and the new equilibrium, resulting from the interplay of the static $\bf h$ and time-dependent $\alpha$. 
With JJ advances and
gate changes exceeding the gigahertz range\cite{Krantz2019:APR}, there is a tantalizing prospect for dynamically controlled CPR by time-dependent SOC.
Unlike assuming a specific relation, $I(\varphi)=I_c \sin(\varphi+\varphi_0)$, 
the CPR can have a more general 
and anharmonic form which should be obtained microscopically. To this end, a single-particle Hamiltonian, 
$H({\bf p})={\bf p}^2/2m^* + {\bf \sigma} \cdot {\bf h} + H_\mathrm{so}({\bf p})$, where $m^*$ is the effective mass, can
be used to solve a BCS model of superconductivity, 
given by the effective Hamiltonian
\begin{equation}
{\cal H}({\bf p}) = \left( \begin{array}{cc}
 H({\bf p}) -\mu \hat{1}& \hat{\Delta} \\
 \hat{\Delta}^\dag & -H^\dagger(-{\bf p}) +\mu \hat{1}
\end{array}\right),
\label{eq:Heff}
\end{equation}
where $\hat{\Delta} $ is a $2\times2$ superconducting gap in spin space\cite{Alidoust2021:PRB}.

After diagonalizing the resulting Bogoliubov–de Gennes equations, ${\cal H}\hat{\psi}=E \hat{\psi}$, where $\hat{\psi}$ is the four-component wave function 
for quasiparticle states with energy $E$, we match the wave functions and generalized velocities 
at interfaces ($x=0, d$), shown in Fig.~\ref{fig:f1}(a). This allows us
to obtain the ground-state JJ energy $E_\mathrm{GS}$, together with the corresponding
CPR, using charge conservation and  the quantum definition of current\cite{Alidoust2021:PRB,Alidoust2021:PRB}.
The CPR is related to the JJ energy:   
$I(\varphi) \propto \partial E_\mathrm{GS}/ \partial \varphi$\cite{Zagoskin:2014}.

Our numerical findings are illustrated for the JJ 
depicted in Fig.~\ref{fig:f1}(a). The normal region ($N$) has a length 
 $L=0.3 \xi_S$ and a width $W=10 L$, such that lengths are normalized by $\xi_\text{S}=\hbar/\sqrt{2m^*\Delta}$,
 where $\Delta$ is the superconducting gap in $S$.
The energies are normalized by $\Delta$ and the supercurrent  
$I_0=2|e\Delta|/\hbar$, where $e$ is the electron charge
and $|e\Delta|/\hbar$ is the maximum supercurrent in a single-channel short $S$-$N$-$S$ JJ\cite{Zagoskin:2014}.
  
\begin{figure}[h]
\includegraphics[width=\columnwidth]{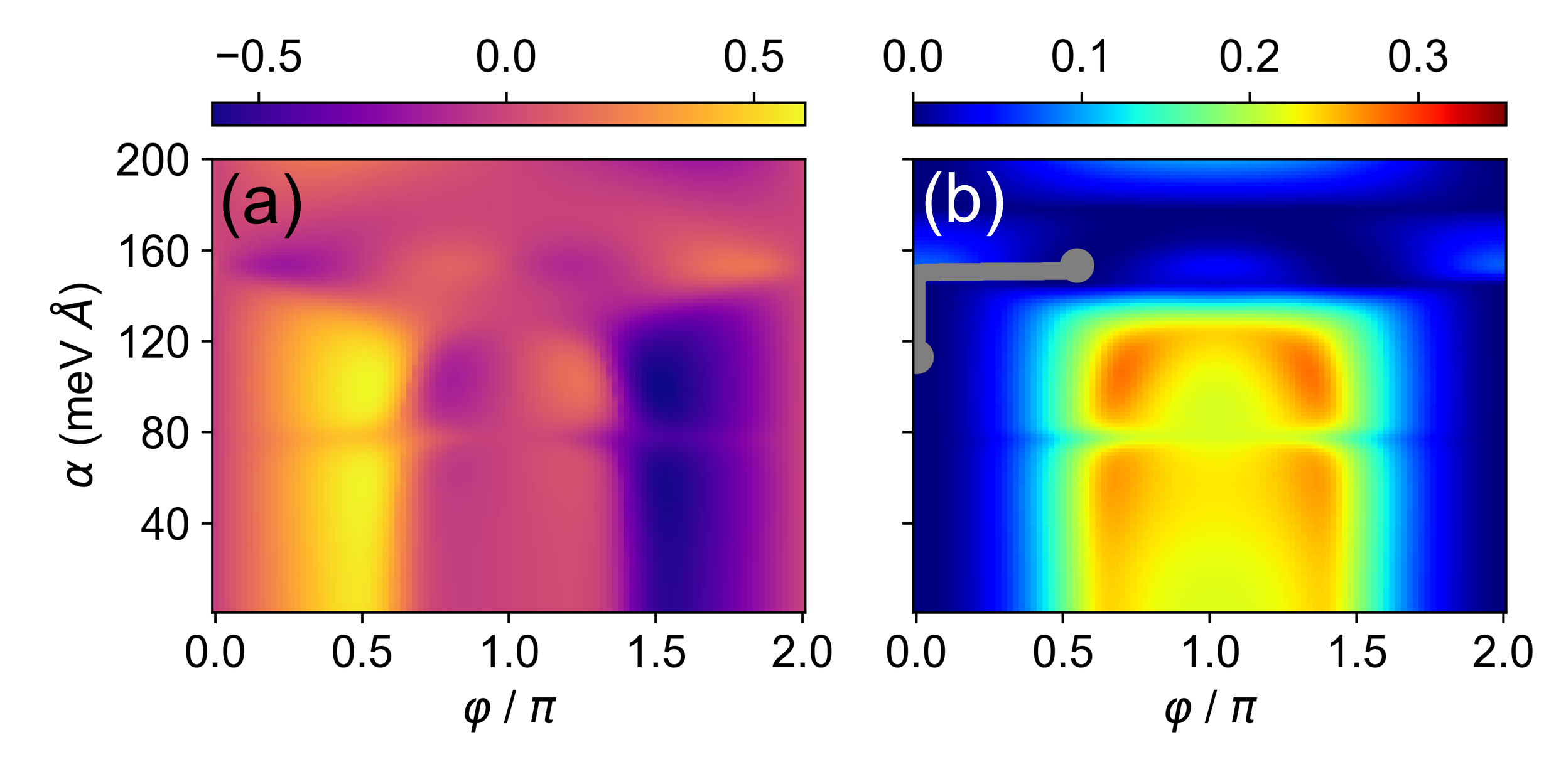}
\caption{(a) The evolution of (a) JJ CPR, normalized by $2|e\Delta|/\hbar$, and (b) the JJ energy, normalized by $\Delta$,
as a function of the phase $\varphi$ and the Rashba SOC, $\alpha$, for chemical potential $\mu = \Delta$ and effective in-plane magnetic field, $h_x=(2/3)\Delta$.  
The gray curve in (b) denotes the JJ transition from  $\varphi=0$ to $\approx \pi/2$ for a linear increase in $\alpha$ from 112 to 152$\,\text{meV\AA}$. 
}
\label{fig:f2}
\end{figure}

To explore the tunability of CPRs and JJ energies with SOC, we focus on the 
parameters for high-quality epitaxial InAs-Al based JJs, 
$\Delta_\text{Al}=0.2\,$meV, with a $g$-factor of 10 for InAs, while its $m^*$ is $0.03\,$ times the electron mass\cite{Mayer2020:NC,Dartiailh2021:PRL}.
In these JJs the gate control of Rashba SOC and thus its magnitude in the range $\alpha \in (0, \, 180 \,\text{meV\AA}$) has been 
demonstrated\cite{Mayer2020:NC,Dartiailh2021:PRL}. In Fig.~\ref{fig:f2}, at  
$h_x=(2/3)\Delta\approx 450\,$mT,  we assume gate control process that primarily changes $\alpha$, not $\mu$. 
Experimentally, this could be realized with dual-gate schemes\cite{vanTuan2019:PRB} to independently tune the carrier density and the electric field, ${\bf E}$.
However,  for a continuous change of $\alpha$, we are unaware that even in a static case the calculated CPR and $E_\mathrm{GS}$ are given.  

In Fig.~\ref{fig:f2}(a), for $\mu=\Delta$, the anharmonic 
CPR changes significantly with $\alpha$. There is a competition between  
$\sin \varphi$ and the next harmonic, $\sin 2\varphi$, resulting in $I(-\varphi)=-I(\varphi)$. However, 
there is no spontaneous current,  $I(\varphi=0)\equiv0$, only $I_c$ reversal with $\alpha$. Such a continuous 
and symmetric 0–$\pi$ transition is well studied without SOC in S/ferromagnet/S JJs due to the changes in the effective magnetization, temperature, or the thickness
of the magnetic region\cite{Kontos2002:PRL,Ryazanov2001:PRL,Bergeret2005:RMP,Eschrig2003:PRL,%
Halterman2015:PRB,Wu2018:PRB,Valls:2022,Yamashita2017:PRA,Yamashita2006:PRB}. 
The corresponding JJ energy landscape in Fig.~\ref{fig:f2}(b), shifted such that its overall  minimum value is 0, 
corroborates this SOC evolution. 
By increasing $\alpha$ from 0 to 200$\,\text{meV\AA}$, the minimum in $E_\mathrm{GS}$ changes from $\varphi=0$
to $\pi$, and then goes back to $0$. A gray trace indicates that by increasing $\alpha$ in a smaller range, 
the JJ minimum can transition from $\varphi=0$ to approximately $\pi/2$.

While we use an exact (complete) CPR with its anharmonicities, their prior descriptions have often relied on an approximate simple harmonic 
expansion ($\sin n\varphi,\, \cos n\varphi$)\cite{Golubov2004:RMP,Kashiwaya2000:RPP}. However, this approach is not very efficient with SOC.
Instead, it is better to use a compact form where only a small number of  terms gives a more accurate description\cite{Alidoust2021:PRB} 
\begin{equation}
I(\varphi, \mu, {\bf h},\alpha) 
\approx \sum_{n=1}^N\sum_{\sigma=\pm}\frac{I_n^\sigma\sin(n\varphi+\varphi_{0n}^\sigma)}{\sqrt{1-\tau_n^\sigma\sin^2(n\varphi/2+\varphi_{0n}^\sigma/2)}},
\label{eq:cpr}
\end{equation}
where  $\tau_n^\sigma$ is the JJ 
transparency for spin channel $\sigma$ and the phase shifts $\varphi_{0n}$ are additional fitting parameters. 
This description includes the anomalous Josephson effect $I(\varphi=0)\neq0$, revisited 
in JJ diode effects\cite{Ando2020:N,Lyu2021:NC,Baumgartner2021:NN,Halterman2021:X,Hu2007:PRL}. 
For a simple picture of a single anomalous phase\cite{Alidoust2021:PRB,Alidoust2020:PRB}.
\begin{equation}
\varphi_0\propto h_y \alpha^3,
\label{eq:anomal}
\end{equation}
therefore vanishing in Fig.~~\ref{fig:f2}, where ${\bf h}=h_x \hat{x}$. 

A quasistatic gate-controlled SOC suggests that more important opportunities are
available using fast gate changes, compatible with the advances in JJ circuits\cite{Krantz2019:APR}. However, the implications of gigahertz changes in
SOC and a different mechanism to drive JJ, as sketched in Fig.~\ref{fig:f1}(b), remain unexplored. To obtain the resulting JJ dynamics
we use Eq.~(\ref{eq:pendulum}) with $i\equiv0$, where the driving arises from $\alpha=\alpha(t)$, viewed as a time-dependent effective ${\bf g'}$. 

Some guidance as to what to expect for JJ dynamics can be given from the InAs-Al samples, where, in addition
to the previous range of $\alpha$, $I_c \sim 4\,\mathrm{\mu A}$, 
$R \sim 100\,\Omega$, and $C \sim15\,\mathrm{fF}$, leading to 
$\omega_p \sim900 \,\mathrm{GHz}$ and the damping $\beta_c \sim 1$, which is also suitable
for the rapid single-flux quantum (RSFQ) applications\cite{Tafuri:2019,Likharev1991:IEEETAS}. We keep $h_x=(2/3)\Delta$.

\begin{figure}[h]
\includegraphics[width=\columnwidth]{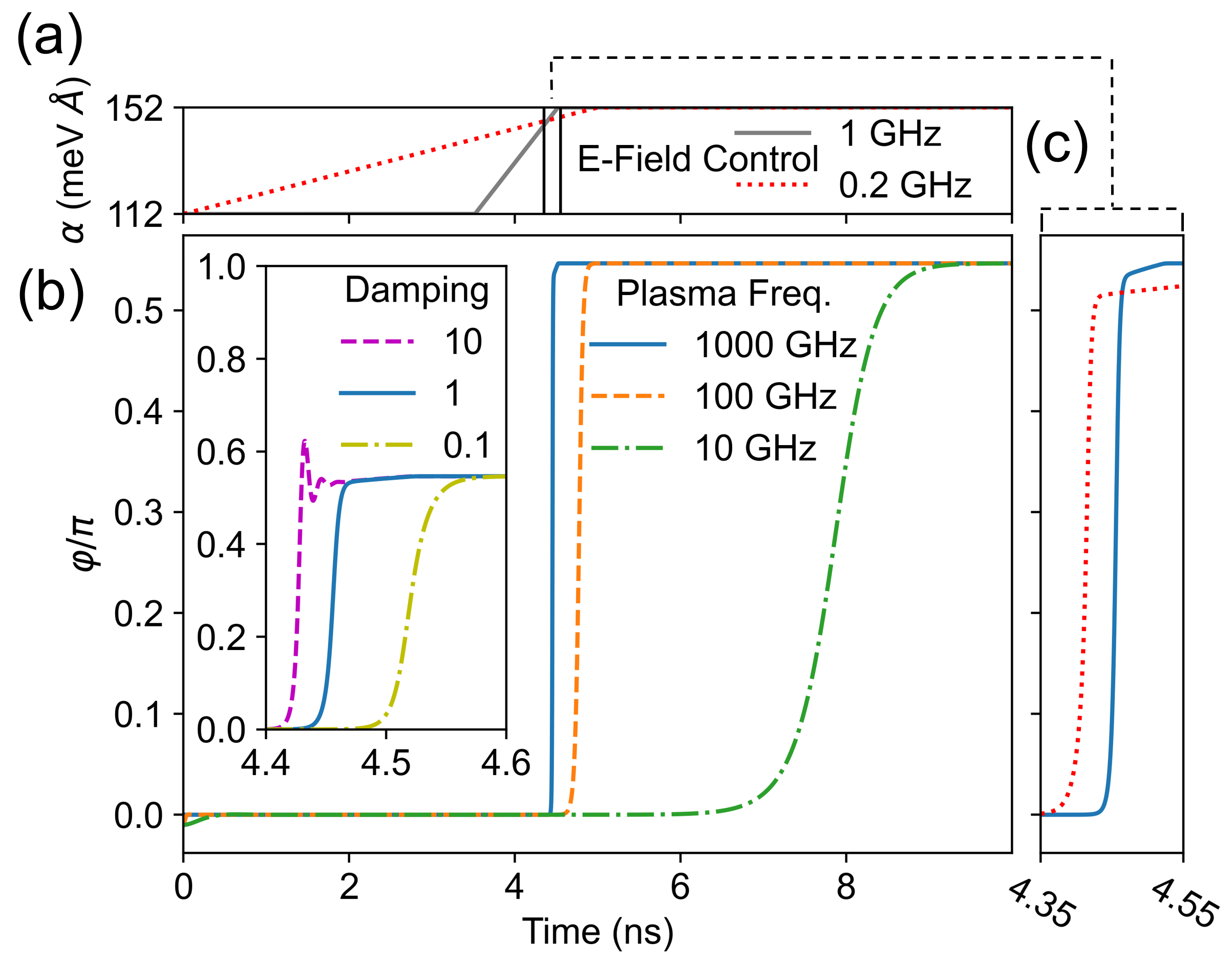}
\caption{(a) A time-dependent Rashba SOC, $\alpha$, controlled by the ${\bf E}$-field, changing at 0.2$\,$GHz and 1$\,$GHz, and also used in (b).
(b) The time-dependent phase for different plasma frequencies, $\omega_p$, at damping, $\beta_c=1$, and for different $\beta_c$ at 
$\omega_p=1000\,$GHz (inset).
(c) An enlarged region for $\varphi=0$ to aporiximately $\pi/2$ transition at 0.2$\,$GHz (1$\,$GHz) dotted (solid) changes in $\alpha$ from (a).
}
\label{fig:f3}
\end{figure}

The JJ dynamics are driven by a simple linear variation of $\alpha(t)$
from the gate-controlled ${\bf E}$, as shown in Fig.~\ref{fig:f3}(a). 
We first consider in Fig.~\ref{fig:f3}(b) the reduction of $\omega_p$, from 1000\,GHz (similar to InAs-Al JJs\cite{Mayer2020:NC}) 
to 10\,GHz (much faster than the $\alpha(t)$ variation), at $\beta_c=1$.  The results reveal a strong delay 
in the onset in the $\varphi=0$ to approximately $\pi/2$ transition, which is
indicated from the static picture in Fig.~\ref{fig:f2}(b). 
Simultaneously, the time for the $\varphi=0$ to approximately $\pi/2$ transition is increased by an order of magnitude. 

We next examine, in the inset of Fig.~\ref{fig:f3}(b), the influence of reducing $\beta_c$ from the underdamped and critical ($\beta_c=10$  and 1) 
to the overdamped ($\beta_c=0.1$) regime, at $\omega_p= 1000\,$GHz.  In addition to 
the phase-oscillation damping, consistent with the pendulum model in Fig.~\ref{fig:f1}(b), 
we also see a delay in the 
$\varphi=0$ to approximately $\pi/2$ transition and its growth, the trends noted from reducing $\omega_p$. 

Finally, in Fig.~\ref{fig:f3}(c), 
the $\varphi=0$ to approximately $\pi/2$ transition occurs first for the slower $\alpha(t)$ variation, 
but takes approximately the same time as the faster gigahertz $\alpha(t)$ variation. This is encouraging for various applications, since
(i) ${\bf E}$ control of SOC allows tailoring of the onset of the transition between different states, (ii) a high-frequency switching 
between different equilibrium states and driving JJs is not limited by the characteristic times for the  ${\bf E}$ variation. 
 $\alpha(t)$ changes at 0.2 GHz give an order-of-magnitude faster transition between the stable phases.

\begin{figure}[h]
\includegraphics[width=\columnwidth]{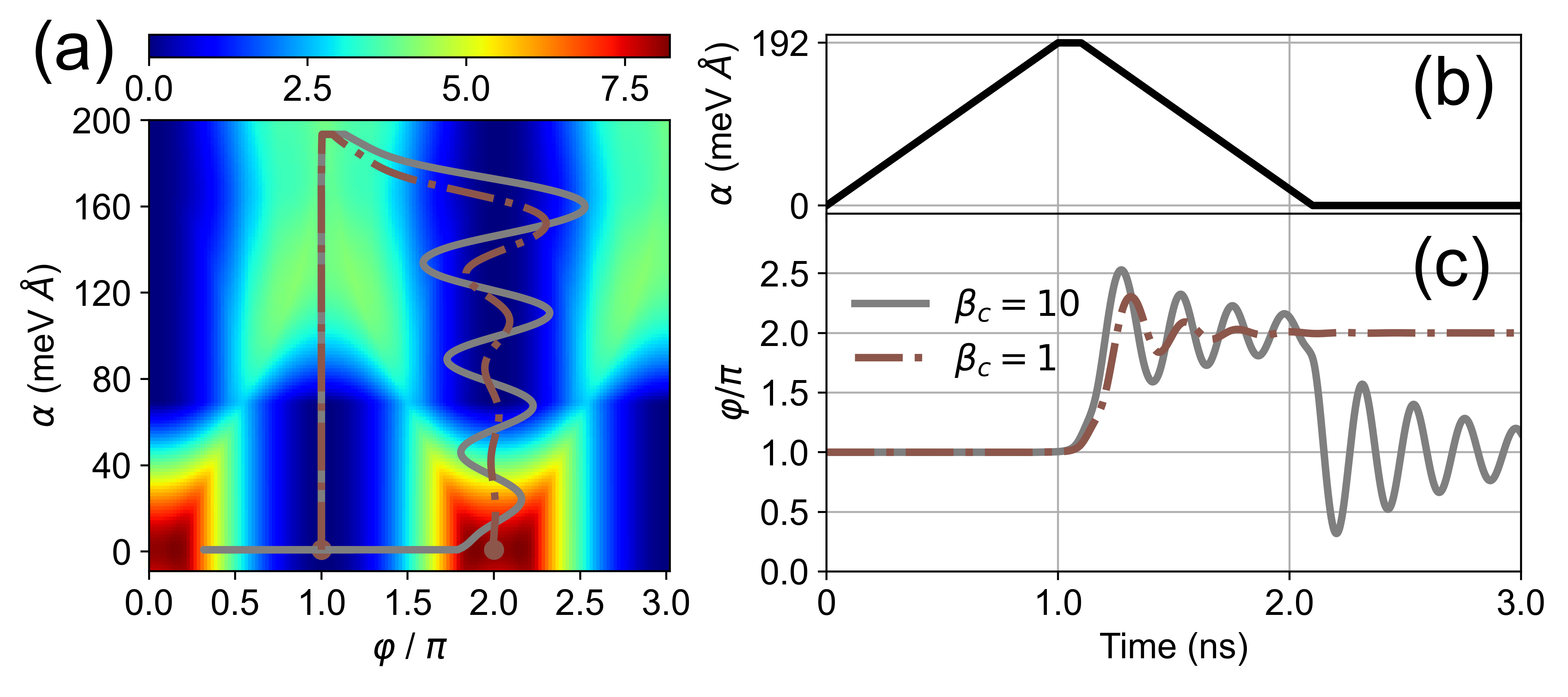}
\caption{(a) The JJ energy evolution with $\varphi$ and $\alpha$ at $\mu=10\Delta$, $h_x=(2/3)\Delta$.
The gray (brown) curve shows the energy variation for $\beta_c=10$ ($\beta_c=1$) starting at $\varphi=\pi$
and $\alpha=0$, for changing $\alpha$, as given in (b). (c) The corresponding 
time-dependent $\varphi$ confirms the decay to different final phase states.
}
\label{fig:f4}
\end{figure}

While the ${\bf E}$ control of $\alpha$ and the 
evolution of the $E_\text{GS}$ minima in Fig.~\ref{fig:f2} largely determine
the JJ dynamics in Fig.~\ref{fig:f3}, it helps to 
identify other opportunities for SOC-driven JJs. 
In Fig.~\ref{fig:f4}, we consider $\omega_p=10\,$GHz and a triangularlike $\alpha(t)$ at $\mu=10\Delta$. For an underdamped
regime, $\beta_c=10$, 
the pendulum analogy from Fig.~\ref{fig:f1}(b) explains the phase evolution of the gray trajectory from Fig.~\ref{fig:f4}(a), also reproduced in Fig.~\ref{fig:f4}(c). 
By increasing $\alpha$ to the maximum 
at $192\,\text{meV\AA}$, the pendulum is at an unstable position and will swing toward the $\varphi=0$ minimum
(equivalently shown as $\varphi=2\pi$), implying that ${\bf g'}$ points vertically down. With small damping 
(gray trajectory), the pendulum passes 
the equilibrium point, even when, with $\alpha <80\,$meV$\text{\AA}$, the equilibrium and the overall minimum shift to $\varphi=\pi$, with ${\bf g'}$ vertically up. 
Eventually, with damping it reaches the $\varphi=\pi$ minimum.

For critical damping,  with the same starting point [see also  Fig.~\ref{fig:f4}(c)], the brown trajectory 
reveals a very different evolution with $\alpha$. 
Instead at the overall $E_\text{GS}$ minimum $\varphi=\pi$,  for $\alpha=0$, the phase is locked at the local minimum $\varphi=0$. With a stronger damping, 
the $\varphi$ oscillations are insufficient to overcome the SOC-dependent barrier which, for $\alpha=0$, separates the local 
minimum at $\varphi=\pi$ from the global one at $\varphi=2\pi$. 
The tunability of the SOC-controlled energy landscape alone does not fully determine the generalized CPRs. 
The influence of the JJ circuit parameters can enable different $\varphi$ transitions.   

\begin{figure}
\includegraphics[width=1.03\columnwidth]{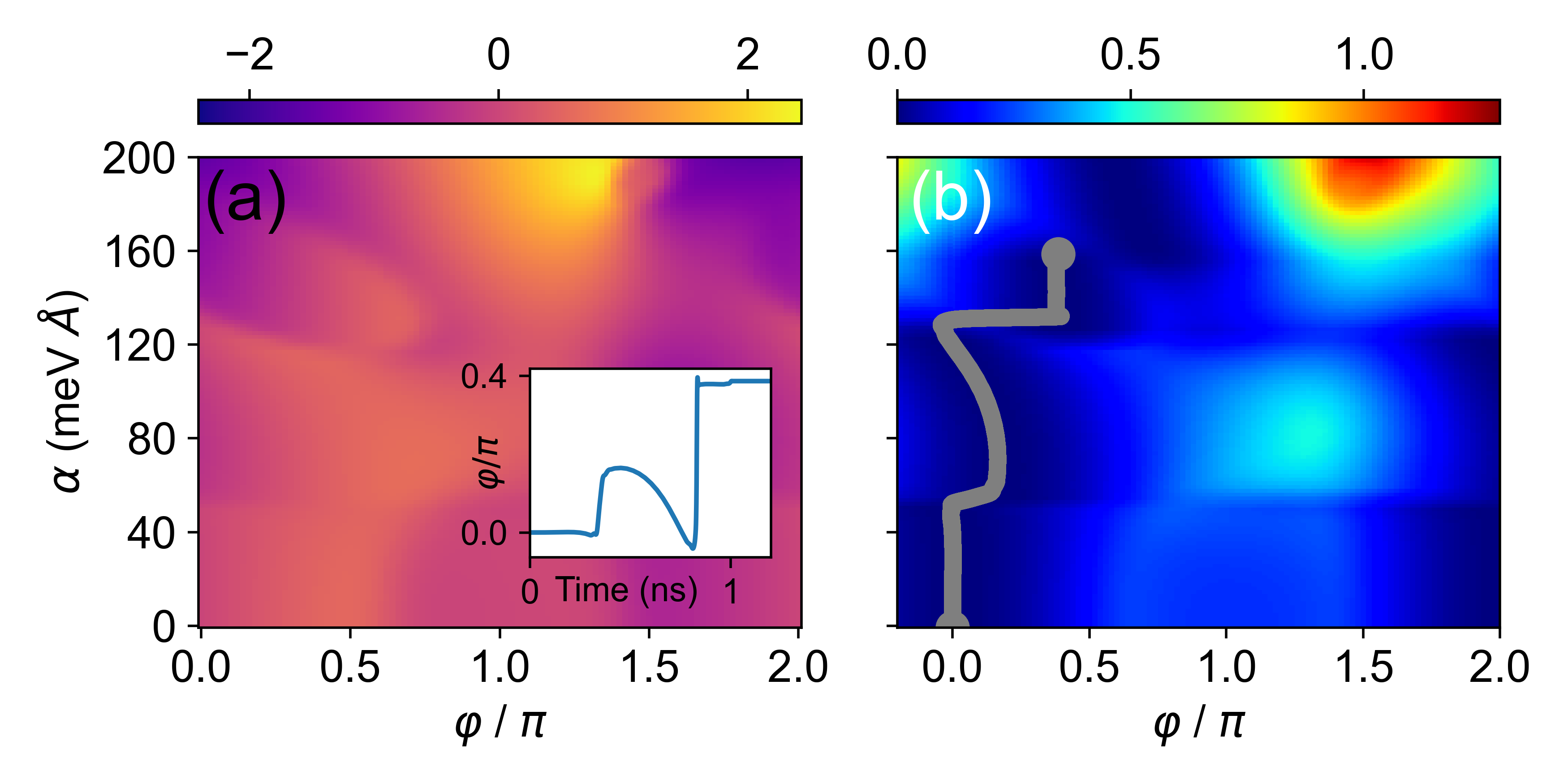}
\caption{The evolution of (a) JJ CPR and (b) the JJ energy with $\varphi$ and $\alpha$, for $\mu = \Delta$ and $h_y=(2/3)\Delta$, 
rotated by $\pi/2$ from Fig.~\ref{fig:f2}. An anharmonic CPR breaks the  $I(-\varphi) = -I(\varphi)$ symmetry in 
(a) and the corresponding anomalous phase, $\varphi_0$,
increases with $\alpha$ in (b). The inset in (a) shows $\varphi(t)$ for $\omega_p=1000\,$GHz, $\beta_c=1$ 
with a linearly increasing $\alpha$ from 0 to 160$\,\text{meV\AA}$ over 1$\,$ns, which is then held at a maximum, with its JJ energy path in (b).
}
\label{fig:f5}
\end{figure}

In the above discussion, the tunability of CPRs and $E_\text{GS}$ does not exploit  
the 
anomalous Josephson 
effect\cite{Reynoso2008:PRL,Buzdin2008:PRL,Konschelle2009:PRL,Sickinger2012:PRL,Xu2022:PRB}, 
which can be understood in analogy to ${\bf g'}$ pointing 
sideways and therefore, breaking the symmetry from Figs.~\ref{fig:f2}-\ref{fig:f4} and $I(-\varphi) \neq -I(\varphi)$. 
This situation can be simply realized by rotating $\bf h$ along the $y$ axis, while we retain all the other parameters from Fig.~\ref{fig:f2}(a).
The resulting CPR in Fig.~\ref{fig:f5}(a) confirms that the JJ supercurrent is driven not only by $\varphi$ but also by $\varphi_0$, which is responsible for the 
stated symmetry breaking and, equivalently, the tilted ${\bf g'}$. 
As for SOC cubic in ${\bf k}$\cite{Alidoust2021:PRB}, there is a 
strong anharmonic behavior and the expected diode effect, where the sign and magnitude of the supercurrent depend on 
 the polarity of the applied bias~\cite{Dartiailh2021:PRL}. 

The implications of the combined broken time-reversal and inversion symmetries, responsible for the anomalous Josephson effect, are
further illustrated in Fig.~5(b), which shows the SOC-tunable $E_\text{GS}$, single valued for the gray path, and leading to the time-dependent diode effect. 
This behavior is qualitatively different from the doubly degenerate $\varphi_0$ state in Fig.~2(b), which results from the second-harmonic generation
in the CPR. 
 
Even with a moderate $h_y\approx 450\,$mT
for InAs based JJs, with increasing $\alpha(t)$ we see an evolution of the single global minimum and thus  
the changes in the corresponding values of $\varphi_0$ from $\varphi=0$ to approximately $3\pi/4$, 
in good agreement with the measured values\cite{Mayer2020:NC}. This suggests that at a larger ${\bf h}$, for example,
in In(As,Sb) with a much larger $g$ factor\cite{Mayer2020:AEM}, it may be possible to fully control the tilt angle of ${\bf g'}$ and simply swap between $0$ 
and $\pi$ states in JJs, further controlling how the JJ dynamics are driven.

The same geometry in Al-InAs JJs at a larger $h_y$ has been experimentally shown to also support topological superconductivity\cite{Dartiailh2021:PRL}. 
This is important for several reasons, beyond hosting Majorana bound states\cite{Aasen2016:PRX}. 
The resulting topological superconductivity is associated with equal-spin $p$-wave superconductivity which could offer
gate-controlled dissipationless spin currents, a key element for superconducting spintronics\cite{Eschrig2015:RPP,Linder2015:NP}. Such spin-triplet supercurrents
could be extended over a long range\cite{Eskilt2019:PRB} and could overcome the usual competition between superconductivity and ferromagnetism. 
A transition to topological superconductivity is accompanied
by an extra phase jump, of approximately $\pi$\cite{Hell2017:PRL,Pientka2017:PRX}.
Such a $\pi$ jump in Al-InAs JJs has been observed at  $h_y\approx600\,$mT\cite{Dartiailh2021:PRL}, 
an effective field about 25 times smaller, than expected for the $0–\pi$ transition 
\begin{equation}
B_{0–\pi}=(\pi/2) \hbar v_\mathrm{F}/(g \mu_\mathrm{B} L),
\label{eq:0pi}
\end{equation}
for a spin-polarized system in the absence of SOC\cite{Yokoyama2014:PRB}, where $v_\mathrm{F}$ is the Fermi velocity,
$\mu_\mathrm{B}$ the Bohr magneton, and $L$ the JJ length.
Therefore,  SOC plays a crucial role in understanding various transitions and, 
at larger $h_y$, the range of an effective  $\varphi_0$ could exceed $2\pi$\cite{Dartiailh2021:PRL} and 
support $2\pi$ pendulum rotation from Fig.~\ref{fig:f1}(b), as used in RSFQ  
logic and memory\cite{Tafuri:2019,Likharev1991:IEEETAS}.
Therefore, in addition to the prospect of fault-tolerant quantum computing, the search for topological superconductivity 
also offers a promising platform for superconducting electronics and spintronics. 

Without previous studies on SOC-driven JJ dynamics, we focus on a simple model and do not consider time-dependent magnetic 
fields\cite{Nashaat2018:PRB} or noise\cite{Massarotti2018:PRB}. A more general description could simultaneously 
include the role of changing $\mu$ and other SOC forms, linear and cubic in ${\bf k}$, 
shown to give different routes to topological 
superconductivity and control of Majorana states\cite{Alidoust2021:PRB,Pekerten2022:PRB,Pakizer2021:PRR,Scharf2019::PRB}.
However, we expect that our focus only on  linearized Rashba SOC,  easily tunable by ${\bf E}$ field\cite{Mayer2020:NC,Dartiailh2021:PRL},
already clarifies its important role in JJ dynamics.  
With changing SOC, there are further opportunities for gate-controlled Majorana states and the probing of their 
non-Abelian statistics\cite{Zhou2022:NC,Paudel2021:PRB} or an added tunability in the implementation of 
superconducting qubits\cite{Krantz2019:APR,Bargerbos2022:X,Hays2021:S}.
This would extend the previously studied qubit tunability by voltage or flux\cite{Krantz2019:APR,Casparis2018:NN} as well as the use of 
$\pi$-phase states for an improved qubit operation\cite{Yamashita2005:PRL,Kawabata2006:PRB}.
 
\acknowledgments
We thank Javad Shabani and Tong
Zhou for valuable discussions. This work is supported
by the National Science Foundation (NSF) Electrical,
Communications and Cyber Systems (ECCS) Grant No.
2130845, the U.S. Office of Naval Research (ONR)
through Grants No. N000141712793 and MURI No.
N000142212764 (D. M. and I. \v{Z}.), and the University at
Buffalo Center for Computational Research

\end{document}